\documentclass[onecolumn,11pt]{IEEEtran}

\usepackage[T1]{fontenc}
\usepackage[utf8]{inputenc}
\usepackage{graphicx}
\usepackage{booktabs}
\usepackage{amsmath}
\usepackage{url}
\usepackage[hidelinks]{hyperref}
\usepackage{xcolor}
\usepackage{framed} 
\usepackage{marvosym}   % \Yinyang, the equal-contribution marker

\graphicspath{{figures/}}

\newenvironment{rqanswer}{\par\medskip\begin{framed}\noindent}{\end{framed}\par\medskip}

\begin{document}

\title{Do Stack Overflow Answer Edits Occur Beyond Java?\\
A Replication on Python and JavaScript}

\author{\IEEEauthorblockN{%
Chaiyong Ragkhitwetsagul\textsuperscript{1}, In-on Wiratsin\textsuperscript{1}, Matheus Paixao\textsuperscript{2}, Denis De Sousa\textsuperscript{2}, Pongpop Lapvikai\textsuperscript{1},
Peter Haddawy\textsuperscript{1}}\\
\bigskip
\IEEEauthorblockA{%
\textsuperscript{1} Faculty of Information and Communication Technology, Mahidol University,
Nakhon Pathom, Thailand\\
\textsuperscript{2} State University of Ceara (UECE), Fortaleza, Brazil\\[4pt]}}

\maketitle

\begin{abstract}
Stack Overflow answers are continually revised by the community, and the edits made to their code snippets are a potential source of improvements for code that has been reused in open-source projects.
A recent empirical study established this for Java, reporting that 16.11\% of accepted Java answers are edited and that the resulting recommendations concentrate in highly popular GitHub projects.
Whether that behaviour is a property of Stack Overflow or a property of the Java community has remained an open question.
We replicate the study on Python and JavaScript, the two most widely used languages alongside Java, applying the same SOTorrent-based extraction pipeline, the same clone search tool, Siamese+, and the same project popularity criteria.
Analysing 840{,}132 accepted Python answers and 1{,}144{,}185 accepted JavaScript answers, we find that 41.25\% and 39.10\% respectively have been edited at least once, roughly two and a half times the Java rate, while the number of revisions per edited answer is almost invariant across the three languages at 2.78, 2.68 and 2.82.
Searching 100 GitHub projects per language, we find that the number of matched answer edits increases monotonically from low- to medium- to high-popularity projects in both languages, from 80 to 156 to 977 for Python and from 32 to 71 to 353 for JavaScript.
The difference is statistically significant for Python but not for JavaScript.
The central findings of the original study therefore generalise beyond Java, with the supply of candidate improvements considerably larger in both replication languages than in the original.
\end{abstract}


\section{Introduction}

Suboptimal code is common in software systems, and developers routinely consult Stack Overflow when writing it.
Because Stack Overflow is collaboratively curated, its answers do not stand still: they are edited after posting, and those edits frequently improve the code they contain, fixing defects, replacing deprecated APIs, or adding the error handling that the first version omitted.
Code copied from an early revision of an answer into a software project does not receive those improvements, so the project silently retains a version of the snippet that the community has since moved on from.
At the same time, code in software projects that are similar to the answer's context may benefit from the improvements, if they can be identified and applied.

A recent empirical study by Wiratsin et al.~\cite{wiratsin2025java} turned this observation into a recommendation pipeline for Java.
The authors extracted the full revision history of Stack Overflow's accepted Java answers from SOTorrent~\cite{baltes2018sotorrent}, indexed every revision of every code snippet in a modified version of the Siamese clone search tool~\cite{ragkhitwetsagul2019siamese}, and searched 10{,}673 GitHub Java projects for code matching an \emph{earlier} revision of an answer.
Where such a match was found, the latest revision of that answer became a candidate improvement.
They reported that 16.11\% of accepted Java answers have more than one revision, that 49.30\% of the resulting recommendations were judged applicable to the projects, and that the recommendations concentrate in the most popular projects.
A subset was submitted as pull requests, of which 21.25\% were accepted by maintainers.

The findings in their study are drawn from Java alone and may not generalise.
That caveat matters more than it might appear, because the mechanism under study depends on community behaviour rather than on anything intrinsic to a programming language.
How often an answer is edited, how large those edits are, and whether the resulting code is reused in projects are all social facts about a particular language community, and there is no reason to assume they transfer.
Related work gives grounds for caution in both directions.
Mondal and Roy~\cite{mondal2025editing} analysed 94{,}994 Python answers and found that edits carry both benefits and costs, improving semantic relevance in 53.3\% of cases while reducing readability in 49.7\%.
Tang and Nadi~\cite{tang2021comment} worked across five languages including Python and JavaScript and found that only 27\% of confirmed comment-edit pairs were potentially useful for code maintenance.
Studies of C/C++~\cite{zhang2022cweaknesses} and C\#~\cite{zuo2025security} report language-specific patterns in how answers evolve.

Java, Python and JavaScript are consistently the most widely used languages in developer surveys and language rankings~\cite{cass2024languages}, and Python and JavaScript differ from Java in ways that could plausibly change the result.
Both are dynamically typed, both have shorter and less ceremonious method bodies, and both sit in ecosystems where library churn is rapid, which may drive more frequent answer revision.
If the findings of the original study hold in these two languages as well, they are much more likely to describe Stack Overflow as a platform than the Java community in particular.

This report therefore replicates the original study on Python and JavaScript.
We follow the same methodology throughout, deliberately changing as little as possible, and we address the following research questions:

\begin{itemize}
\item \textbf{RQ1:} \emph{To what extent are accepted Python and JavaScript answers on Stack Overflow edited?}
As in the original study, this establishes the supply of potential code improvements.
The proportion of accepted answers that carry an edit bounds how much of the platform's code could be improved by adopting a later revision. Thus, if that proportion is much lower outside Java the approach has correspondingly less to work with.
Establishing it for two further languages also shows whether the findings from Java are representative of Stack Overflow or particular to one community.
\item \textbf{RQ2:} \emph{How are the Stack Overflow answer edits matched in open-source projects distributed across projects of different popularity?}
An edited answer is only actionable if the code it supersedes is present in real projects, so this question investigates whether the matches observed for Java also arise in Python and JavaScript.
The distribution across popularity groups is of particular interest because the original study reported that the most popular projects, rather than the least, contain the most matched edits, and a replication offers the first evidence of whether that pattern extends beyond Java.
\end{itemize}

Our RQ2 is narrower than the original.
The original study classified each matched recommendation manually into \emph{Fixing Bug} and \emph{Improving Code} categories and reported an applicability rate.
We perform no manual classification, so we report how many answer edits were matched in each popularity group but not what kind of improvement each one represents.
We also do not replicate the original RQ3, an objective static-analysis comparison of original and revised snippets, nor the pull-request submission phase.

The contributions of this report are as follows.

\begin{enumerate}
\item A replication of the answer-edit analysis on 840{,}132 accepted Python answers and 1{,}144{,}185 accepted JavaScript answers, giving the first three-language comparison of Stack Overflow answer editing under a single, identical operationalisation.
\item Findings and implications based on clone search of 100 GitHub projects per language with Siamese+, showing how the matched answer edits distribute across project popularity groups, and comparisons to the original study.
\end{enumerate}

\section{Methodology}

We follow the methodology of the original study~\cite{wiratsin2025java}, reusing its pipeline and its parameters wherever possible so that any difference in the results can be attributed to the language rather than to the method.
This section describes the shared pipeline first, then the details specific to each language.

\subsection{Datasets}

Stack Overflow data come from SOTorrent~\cite{baltes2018sotorrent}, which augments the official data dump with the version history of every post, retrievable at the level of the individual post block.
We used a local MySQL copy of the same release as the original study, version 2020-12-31, which contains 51{,}296{,}931 posts and 81{,}536{,}422 post versions.
Using the same release keeps the temporal coverage of the three studies identical.

Tags in Stack Overflow are attached to questions rather than to answers, so answers of a given language are identified through their parent question, by joining \texttt{Posts} to \texttt{PostTags} tables and resolving the language tag to its \texttt{Tags.Id}.
An answer enters our set when it is both \emph{accepted}, that is its \texttt{Id} equals the parent question's \texttt{AcceptedAnswerId}, and \emph{revised}, that is it has at least one \texttt{PostVersion} row whose \texttt{PredPostHistoryId} is not null, which marks a version that has a predecessor and therefore an edit.

For each such answer we extracted the code snippets from \texttt{PostBlockVersion} table, where code snippets are distinguished from text blocks.
Answers containing more than one code snippet were neither merged nor reduced to a single one.
Each code snippet is tracked separately by its \texttt{LocalId}, which gives the code snippet's position within a revision and remains stable across the revisions in which that code snippet exists, so an answer with two code snippets contributes two independent edit histories.
For every \texttt{(PostId, LocalId)} pair we wrote out each revision of the code snippet as its own file, along with two distinguished versions: the \emph{latest}, the row with \texttt{MostRecentVersion = 1}, and the \emph{original}, the root of the code snippet's edit chain given by \texttt{RootPostBlockVersionId}.
Taking the original from the root of the chain rather than from the answer's first revision matters because an edit chain can break and restart when a code snippet is removed and later re-added; in that case the earliest ancestor of the current chain, rather than the very first version of the post, is the snippet from which the latest version was actually derived.

\subsection{Measuring the size of an edit}

To quantify how much a code snippet changed, we computed the Levenshtein distance~\cite{levenshtein1966binary} between the original and the latest version of each code snippet, that is the minimum number of single-character edits needed to transform one into the other.
This is the same measure the original study used.
We report the distribution over all pairs, and, because a large fraction of pairs turn out to be identical, we also report the distribution restricted to the pairs that actually changed.

\subsection{GitHub project selection}

The original study collected projects with GitHub Search~\cite{dabic2021ghs} using the filters \texttt{Language}, \texttt{Exclude Forks}, \texttt{Has Open Issues} and \texttt{Has Open Pull Requests}, then grouped them by popularity.
Three metrics were used, the numbers of stars, watchers and forks.
Each metric's distribution was divided into quartiles, and a project was assigned to the low-popularity group when all three of its metric values fell in the first quartile, to the high-popularity group when all three fell in the fourth quartile, and to the medium-popularity group when all three fell between the first and third quartiles.
Projects landing in different quartiles for different metrics were excluded, which gives a clean separation between the groups.

Unfortunately, GitHub Search is no longer working at the time of this replication, so we cannot retrieve the set of all projects in a language and compute quartiles from it. We resorted to using GitHub REST API v3~\cite{githubapi} to retrieve the Python and JavaScript projects based on their stars and forks counts (watchers cannot be queried using the API due to the changes in the API~\cite{githubwatching}: the \texttt{watchers} and \texttt{watchers\_count} fields now report the number of users who have starred a repository rather than the number watching it, and the subscriber endpoints that carry the true watcher count are restricted).
Due to this missing watcher information, we inherit the resulting thresholds based only on stars and forks: 10 to 25 stars and 0 to 14 forks for the low-popularity group, 26 to 271 stars and 15 to 105 forks for the medium-popularity group, and 272 or more stars and 106 or more forks for the high-popularity group.
Reusing the Java thresholds rather than recomputing quartiles per language keeps the three studies directly comparable, at the cost of the groups no longer being quartiles of the Python or JavaScript project populations themselves.
We discuss this threat to validity in Section~\ref{sec:threats}.

We searched 100 projects per language, 33 low-popularity, 33 medium-popularity and 34 high-popularity.
This is a balanced sample and is far smaller than the 10{,}673 projects of the original study, a difference that has consequences for the statistical power of our RQ2 analysis and for the comparability of the absolute counts.
Both are discussed in the Discussion and Threats to Validity (Sections~\ref{sec:discussion} and~\ref{sec:threats}). 

\subsection{Clone search with Siamese+}

To locate adoptions of Stack Overflow code in the projects we used Siamese+,\footnote{\url{https://github.com/cragkhit/Matcha}} the modified clone search tool introduced by the original study.
Siamese+ extends Siamese~\cite{ragkhitwetsagul2019siamese}, a scalable clone search tool that transforms code into multiple representations and is accurate for Type-1 (exact-copy code) to Type-3 clones (copied code with modifications), with three modules that the original study added for this purpose.

\begin{itemize}
\item \emph{Boiler-plate code filter}, which excludes trivial queries such as getters, setters and equality methods, encoded as regular expressions following a previous study~\cite{ragkhitwetsagul2021toxic}, so that trivial recommendations are not returned. We added the list of boiler plate code patterns for Python and JavaScript, following the same approach as the original study.
\item \emph{Multiple code revision search}, which indexes every revision of every code snippet, naming each snippet by concatenating its \texttt{PostId}, \texttt{LocalId} and \texttt{HistoryId} and marking the first and last revisions as \texttt{original} and \texttt{latest}, so that a match can be recognised as belonging to an earlier revision by inspecting its file name.
\item \emph{Latest code revision retrieval}, which returns the latest revision of a matched code snippet when the match itself was to an older revision. 
\end{itemize}

As used by Wiratsin et al.~\cite{wiratsin2025java}, Siamese+ parsed and indexed Java only, which is what confined that study to a single language.
After the implementation of Siamese+, the original Siamese has since been extended with language support for Python and JavaScript, and it is that extension which makes the present replication possible.
For JavaScript, the support follows the extension of Siamese to JavaScript reported by Misu and Satter~\cite{misu2022javascript}, who applied it to the analysis of JavaScript online code clones.
The extension adds the parsing and tokenisation needed to read the two languages and to segment their source into the method-level units the search operates on. It does not alter the matching algorithm or the search parameters, which are discussed next.
For Python, the creator of Siamese has added the parsing and tokenisation needed to read Python source and segmentation into method-level units some time after its first support for Java. 
We incorporated that extension into Siamese+ for this study, again without changing the matching algorithm or the search parameters.

The original study tuned Siamese in a preparatory phase using Grid Search over 3{,}073 configurations, maximising mean reciprocal rank against a ground truth of clone pairs between Stack Overflow and the Qualitas corpus~\cite{tempero2010qualitas}, and obtained a best configuration with a clone size of 6 lines, n-gram sizes of 1 and 4, query reduction thresholds of 9, 6, 5 and 9, and similarity thresholds of 50\%, 60\%, 70\% and 80\%, scoring an MRR of 0.782.
\textbf{We reuse that configuration unchanged for both Python and JavaScript.}
No language-specific retuning was performed.
This is a deliberate choice: it keeps the three languages comparable and faithful to the original method, but it means the tool is operating with parameters optimised for Java code, which we treat as a threat to validity in Section~\ref{sec:threats}.

\subsection{Counting answer edits in the search results}
\label{sec:counting}

Siamese+ returns, for each method in a project, the ranked list of Stack Overflow snippets it matched.
Because the index holds every revision of every code snippet, one project method typically matches several revisions of the same code snippet, so we fold those together and take the code snippet, identified by its \texttt{(PostId, LocalId)} pair, as the unit of counting.
We report three quantities per popularity group:

\begin{itemize}
\item \textbf{candidate pairs}, the number of result lines with at least one match, which is the closest analogue of the 793 pairs reported by the original study;
\item \textbf{code snippets matched}, the number of distinct code snippets matched anywhere in the group;
\item \textbf{answer edits}, the number of matched code snippets whose content actually changed between the original and the latest revision, verified against the Levenshtein distances.
\end{itemize}

The last of these is our headline figure.
A matched code snippet with a distance of zero was matched but never edited, so it carries no improvement that could be recommended, and counting it would overstate the supply.
Group totals are de-duplicated, so a code snippet matched in three projects of the same group is counted once. Per-project figures are also counted once per project.

Following the original study, we test whether the counts differ across the three groups.
The null hypothesis is that there is no statistically significant difference in the number of matched answer edits across the low-, medium- and high-popularity groups.
We first check normality with the Shapiro-Wilk test~\cite{shapiro1965normality} and, where it is rejected, use the non-parametric Kruskal-Wallis test~\cite{kruskal1952ranks} on the per-project counts, exactly as the original study did.

\subsection{Language-specific details}

The pipeline is language-agnostic apart from three points.

\textbf{Python.}
Answers were selected through the \texttt{python} tag, which carries 1{,}597{,}896 questions in the dataset.
Extracted snippets were written with the \texttt{.py} extension.

\textbf{JavaScript.}
Answers were selected through the \texttt{javascript} tag, which carries 2{,}130{,}783 questions.
Snippets were written with the \texttt{.js} extension.

\textbf{Java.}
All Java figures quoted in this report are taken from the original study~\cite{wiratsin2025java} rather than recomputed, so that the comparison is against the existing result.

\section{Results}

We report RQ1 and RQ2 for Python and then for JavaScript.
Java figures from the original study are given alongside for reference throughout.

\subsection{Python}
\label{sec:results-python}

\subsubsection{RQ1: To what extent are accepted Python answers edited?}

The \texttt{python} tag carries 1{,}597{,}896 questions, of which 840{,}132 have an accepted answer.
Those accepted answers contain 1{,}190{,}274 code snippets in their latest version, an average of 1.42 snippets per accepted answer with a standard deviation of 1.32.
Of the 840{,}132 accepted answers, 346{,}535 have been edited at least once, that is \textbf{41.25\%}.
Of those, 305{,}768, or 88.24\%, contain at least one code snippet and form the set used for all subsequent analysis; they amount to 36.40\% of all accepted answers.
Across all revisions of that set there are 1{,}640{,}080 code snippets, which were indexed in Siamese+.

\begin{figure}[!t]
\centering
\includegraphics[width=0.6\linewidth]{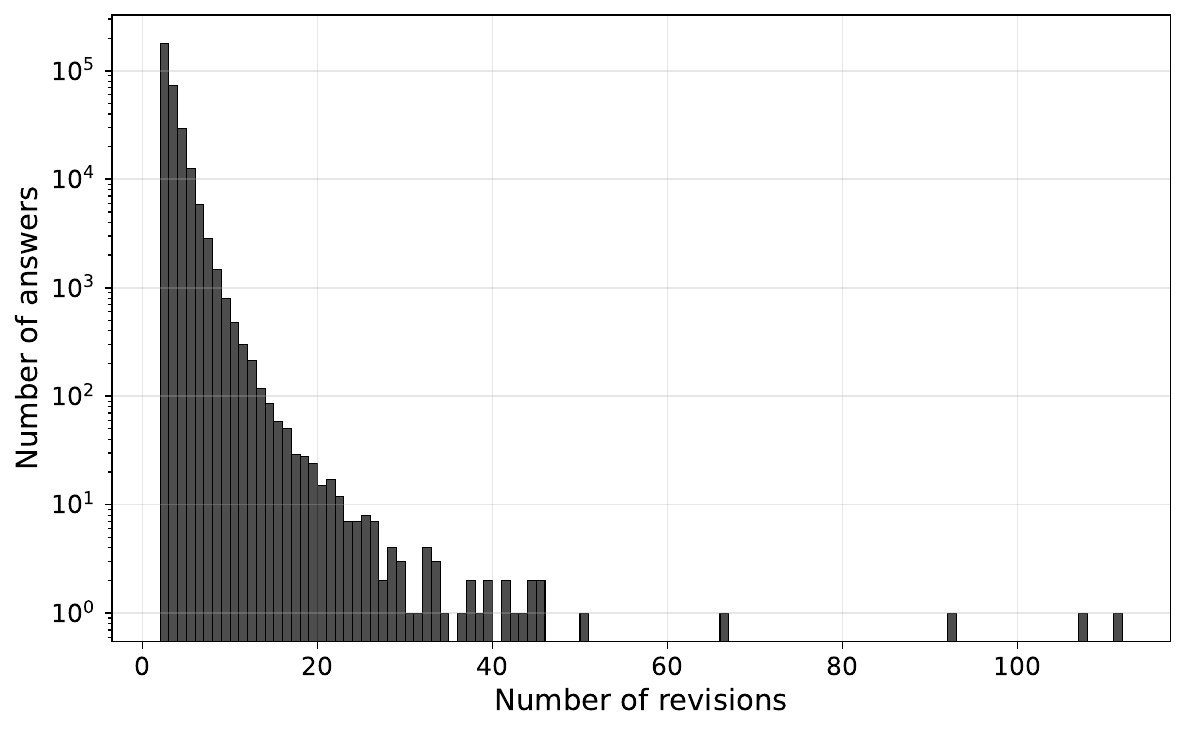}
\caption{Distribution of the number of revisions in the filtered Python accepted answers.
The vertical axis is logarithmic.}
\label{fig:python-revisions}
\end{figure}

The average number of revisions per edited answer is 2.78, with a median of 2.0 and a standard deviation of 1.43.
The distribution is given in Figure~\ref{fig:python-revisions}.
The most revised answer in our set is post 60662471,\footnote{\url{https://stackoverflow.com/posts/60662471/revisions}} with 111 revisions made between 12 March 2020 and 17 March 2020, a burst of editing spanning just over four days. 

To gauge the size of the code changes we computed the Levenshtein distance between the original and the latest version of each code snippet, over 637{,}923 pairs.
The minimum is 0, the maximum 28{,}500, the mean 49.66 and the standard deviation 268.53.
The median, however, is 0: \textbf{438{,}152 pairs, or 68.68\% of the total, are byte-identical}.
In other words, in around two thirds of cases the answer was edited but that particular code snippet was not.
Restricted to the 199{,}771 pairs that did change, the mean distance is 158.57 and the median 43.
The quartiles of the full distribution are 0, 0 and 7, with the 90th, 95th and 99th percentiles at 103, 244 and 866 characters.
Most code edits are therefore small, but a long tail of very large edits remains, as Figure~\ref{fig:boxplots} shows.

Aggregating to the level of the answer, 303{,}271 answers have at least one original-latest code snippet pair, an average of 2.10 code snippets per edited answer, and 163{,}360 of them contain at least one code snippet that actually changed.
That is 53.87\% of the code-bearing edited answers and 19.44\% of all accepted Python answers.

\begin{rqanswer}
\textbf{Answer to RQ1 (Python):} 41.25\% of Stack Overflow Python accepted answers have more than one revision, with the average number of revisions per answer being 2.78.
The average code edit size between the original and the latest revision is 49.66 characters, although 68.68\% of code snippets are left untouched by the edit.
\end{rqanswer}

\subsubsection{RQ2: How are the matched answer edits distributed across project
popularity?}

\begin{figure}[!t]
\centering
\includegraphics[width=0.92\linewidth]{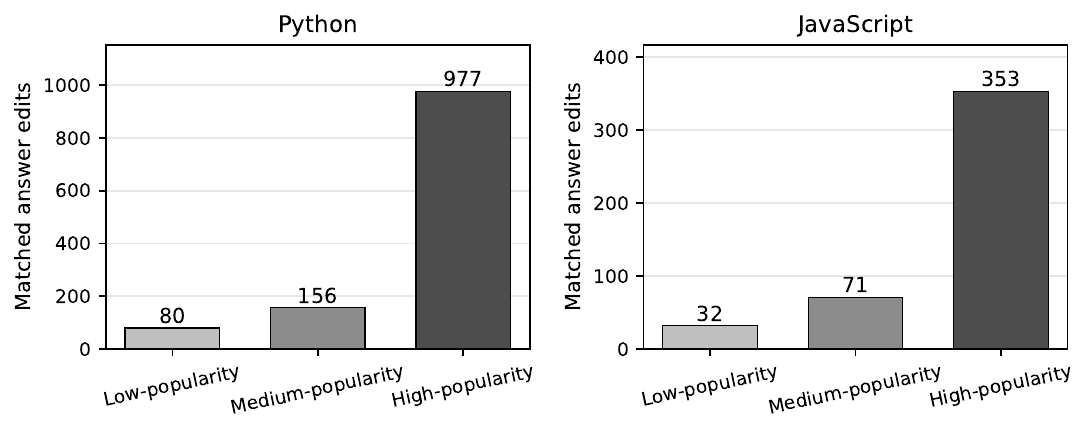}
\caption{Matched Stack Overflow answer edits grouped by GitHub project popularity, for
Python and JavaScript.
Note that the vertical scales differ between the two panels.}
\label{fig:rq2}
\end{figure}

Siamese+ was run with the tuned configuration over the 100 selected Python projects.
It returned 754 candidate pairs, covering 1{,}970 distinct code snippets, of which 1{,}080 correspond to code snippets that were genuinely edited.
The distribution across the three popularity groups is given in Table~\ref{tab:rq2-python} and plotted in Figure~\ref{fig:rq2}.

\begin{table}[!t]
\caption{Matched Stack Overflow answer edits in the Python GitHub projects, grouped by
project popularity.
Max and Avg are the maximum and average number of matched answer edits per project.}
\label{tab:rq2-python}
\centering
\begin{tabular}{lrrrrrrr}
\toprule
Group & Projects & With matches & Candidate pairs & Snippets & Answer edits & Max & Avg \\
\midrule
Low-popularity    & 33 &  7 &  31 &   149 &  80 &  54 &  2.4242 \\
Medium-popularity & 33 & 16 & 111 &   304 & 156 &  64 &  5.6364 \\
High-popularity   & 34 & 21 & 612 & 1{,}767 & 977 & 777 & 36.8824 \\
\midrule
Total & 100 & 44 & 754 & 1{,}970 & 1{,}080 & & \\
\bottomrule
\end{tabular}
\end{table}

The counts increase monotonically with popularity on every measure.
The number of projects containing at least one match rises from 7 to 16 to 21, the candidate pairs from 31 to 111 to 612, and the matched answer edits from 80 to 156 to 977.
The average number of matched edits per project rises by more than an order of magnitude across the three groups, from 2.42 to 5.64 to 36.88, and the single most affected project, in the high-popularity group, matched 777 edited code snippets.

The Shapiro-Wilk test rejects normality in all three groups, with $p < 10^{-8}$ throughout, so we used the Kruskal-Wallis test on the per-project counts.
It reports a statistic of $H = 12.599$ with 2 degrees of freedom and $p = 0.001837$.
At $\alpha = 0.05$ we reject the null hypothesis and conclude that there is a statistically significant difference in the number of matched answer edits across the three project groups.

\begin{rqanswer}
\textbf{Answer to RQ2 (Python):} Siamese+ matched 1{,}080 edited Stack Overflow code snippets across 100 Python projects.
The number of matched answer edits increases with project popularity, from 80 in the low-popularity group to 156 in the medium-popularity group and 977 in the high-popularity group, and the difference across the groups is statistically significant ($p = 0.0018$).
\end{rqanswer}

\subsection{JavaScript}
\label{sec:results-javascript}

\subsubsection{RQ1: To what extent are accepted JavaScript answers edited?}

The \texttt{javascript} tag carries 2{,}130{,}783 questions, of which 1{,}144{,}185 have an accepted answer, making it the largest of the three answer sets by some margin.
Those answers contain 1{,}591{,}776 code snippets in their latest version, an average of 1.39 per answer with a standard deviation of 1.24.
Of the accepted answers, 447{,}379 have been edited at least once, that is \textbf{39.10\%}.
Of these, 391{,}264, or 87.45\%, contain at least one code snippet, amounting to 34.20\% of all accepted answers.
Across all revisions there are 1{,}998{,}016 code snippets, the largest index of the three studies.

\begin{figure}[!t]
\centering
\includegraphics[width=0.6\linewidth]{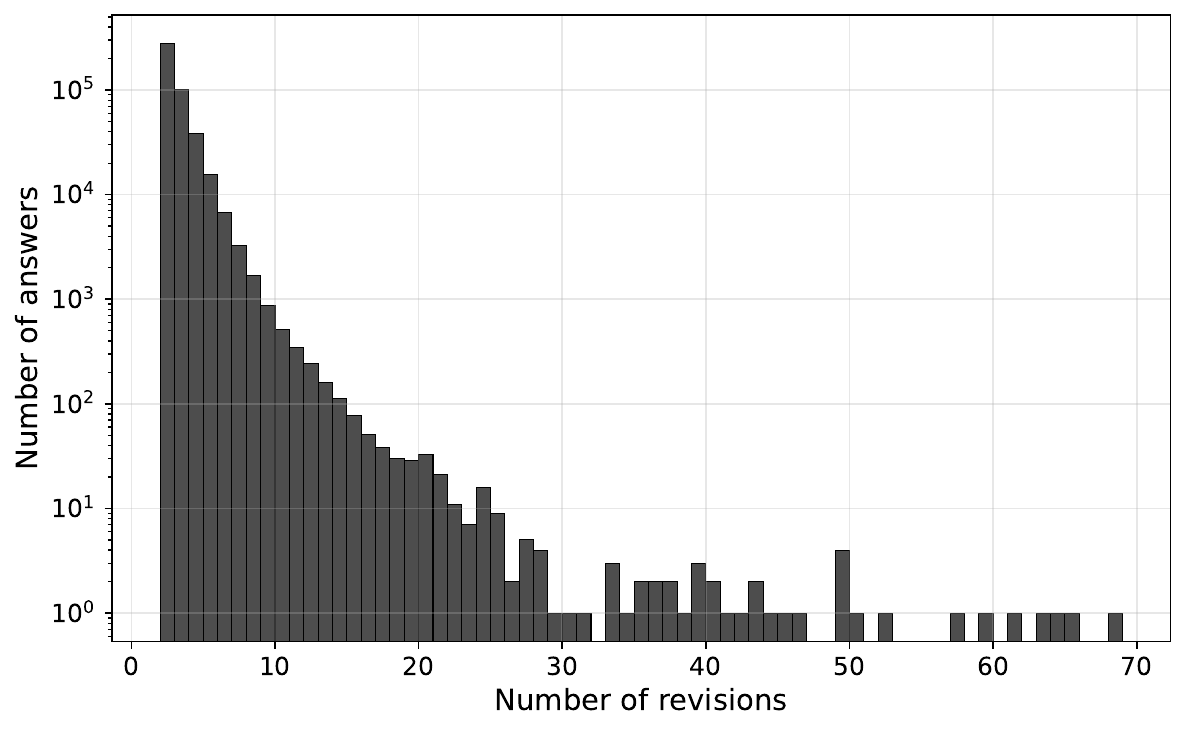}
\caption{Distribution of the number of revisions in the filtered JavaScript accepted
answers.
The vertical axis is logarithmic.}
\label{fig:javascript-revisions}
\end{figure}

The average number of revisions per edited answer is 2.68, with a median of 2.0 and a standard deviation of 1.31.
The distribution is shown in Figure~\ref{fig:javascript-revisions}.
The most revised answer is post 9550412,\footnote{\url{https://stackoverflow.com/posts/9550412/revisions}} with 68 revisions, but unlike the Python maximum these are spread over eight years and eight months, from 3 March 2012 to 26 November 2020, which is sustained maintenance rather than a burst.\footnote{This reported result is based on the SOTorrent snapshot version 2020-12-31. Nonetheless, after that, the answer had been continually edited. The current number of revisions, at the time of writing, is 85, spread over eleven years and two months, from 3 March 2012 to 19 May 2023.}

The Levenshtein distances were computed over 793{,}362 original-latest pairs.
The minimum is 0, the maximum 20{,}690, the mean 48.66 and the standard deviation 228.44, with a median of 0.
As with Python, the great majority of code snippets are untouched: 537{,}142 pairs, or 67.70\%, are byte-identical.
Restricted to the pairs that changed, the mean is 150.68 and the median 48.
The quartiles are 0, 0 and 8, with the 90th, 95th and 99th percentiles at 115, 253 and 798 characters.

At the level of the answer, 387{,}883 answers have at least one original-latest pair, an average of 2.05 code snippets per edited answer, and 207{,}945 contain at least one code snippet that changed, which is 53.61\% of the code-bearing edited answers and 18.17\% of all accepted JavaScript answers.

\begin{figure}[!t]
\centering
\includegraphics[width=0.6\linewidth]{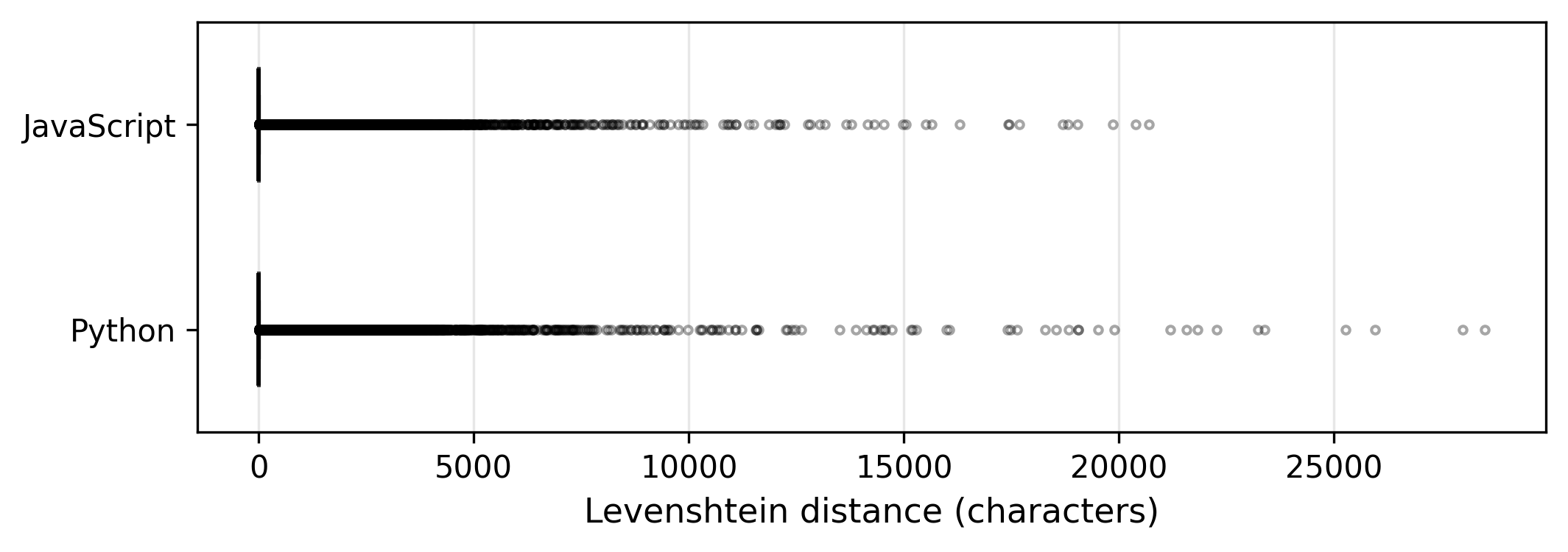}\\[6pt]
\includegraphics[width=0.6\linewidth]{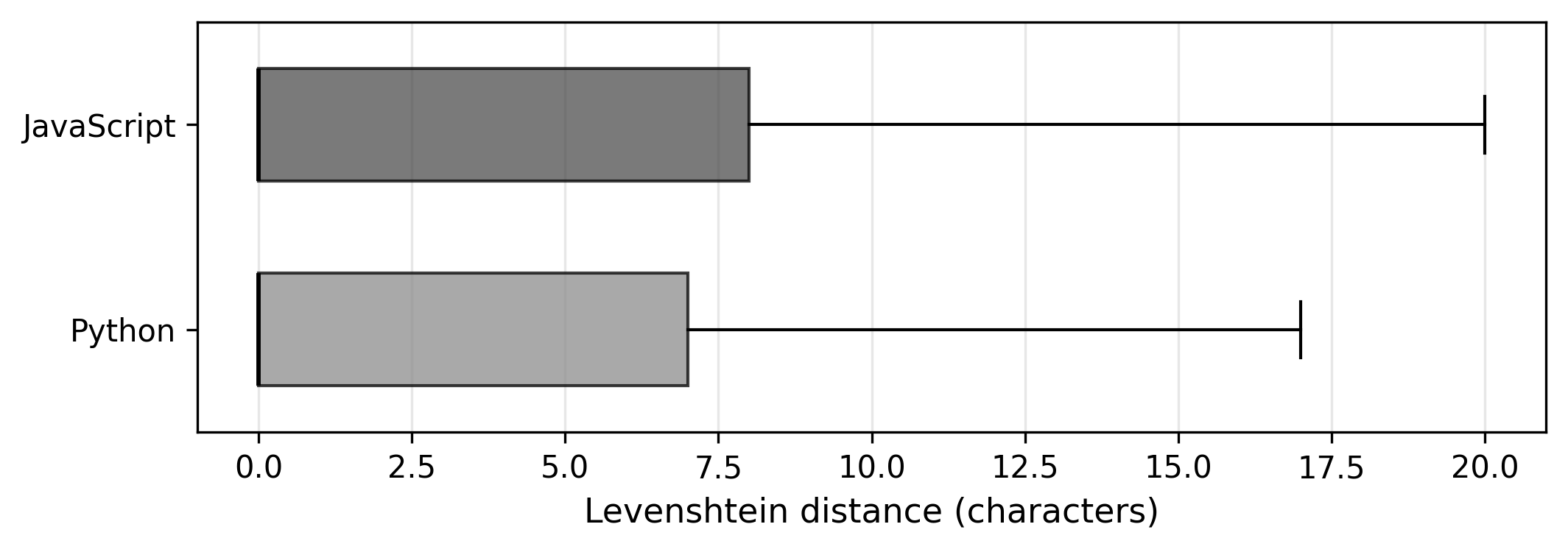}
\caption{Distribution of the Levenshtein distance between the original and the latest
Stack Overflow code answers, for Python and JavaScript.
The upper pair shows the full distribution, in which the long tail of very large edits dominates the axis.
The lower pair hides the outliers. Note the change of scale between the two.}
\label{fig:boxplots}
\end{figure}

The agreement between the two replication languages on these distance statistics is close enough to be worth noting explicitly.
The proportion of unchanged code snippets differs by less than one percentage point, 67.70\% against 68.68\%, the mean distances by one character, and the quantiles are near-identical at every level.
Figure~\ref{fig:boxplots} shows the two distributions together.

\begin{rqanswer}
\textbf{Answer to RQ1 (JavaScript):} 39.10\% of Stack Overflow JavaScript accepted answers have more than one revision, with the average number of revisions per answer being 2.68.
The average code edit size between the original and the latest revision is 48.66 characters, although 67.70\% of code snippets are left untouched by the edit.
\end{rqanswer}

\subsubsection{RQ2: How are the matched answer edits distributed across project
popularity?}

Over the 100 selected JavaScript projects, Siamese+ returned 546 candidate pairs covering 970 distinct code snippets, of which 449 correspond to genuinely edited code snippets.
Table~\ref{tab:rq2-javascript} gives the breakdown by popularity group.

\begin{table}[!t]
\caption{Matched Stack Overflow answer edits in the JavaScript GitHub projects, grouped by
project popularity.
Max and Avg are the maximum and average number of matched answer edits per project.}
\label{tab:rq2-javascript}
\centering
\begin{tabular}{lrrrrrrr}
\toprule
Group & Projects & With matches & Candidate pairs & Snippets & Answer edits & Max & Avg \\
\midrule
Low-popularity    & 33 & 10 &  70 &  87 &  32 &  16 &  1.0303 \\
Medium-popularity & 33 & 10 &  62 & 164 &  71 &  28 &  2.1818 \\
High-popularity   & 34 & 16 & 414 & 749 & 353 & 121 & 11.7941 \\
\midrule
Total & 100 & 36 & 546 & 970 & 449 & & \\
\bottomrule
\end{tabular}
\end{table}

The same monotonic pattern appears.
Matched answer edits rise from 32 to 71 to 353, and the per-project average from 1.03 to 2.18 to 11.79, an elevenfold increase from the lowest to the highest group.
The number of projects containing at least one match is equal in the low- and medium-popularity groups, at 10 each, and rises to 16 in the high-popularity group.

The Shapiro-Wilk test again rejects normality in all three groups, with $p < 10^{-9}$ throughout.
The Kruskal-Wallis test reports $H = 3.051$ with 2 degrees of freedom and $p = 0.217479$.
At $\alpha = 0.05$ we \emph{retain} the null hypothesis: although the counts clearly increase with popularity, the difference between the three groups is not statistically significant for JavaScript.
Pairwise Mann-Whitney tests, reported here only as exploratory context, likewise show no significant separation between any pair of groups.

The most plausible explanation is limited statistical power rather than an absence of effect.
Each group contains only 33 or 34 projects, against the thousands in the original study, and the per-project counts are extremely skewed: the median project in every JavaScript group matched no edits at all, so the test is comparing three samples that are mostly zeros with a small number of large values.
The JavaScript counts are also roughly a third of the Python ones in every group, which compounds the problem.
We return to this in Section~\ref{sec:discussion}.

\begin{rqanswer}
\textbf{Answer to RQ2 (JavaScript):} Siamese+ matched 449 edited Stack Overflow code snippets across 100 JavaScript projects.
The number of matched answer edits increases with project popularity, from 32 in the low-popularity group to 71 in the medium-popularity group and 353 in the high-popularity group, but with only 33 or 34 projects per group the difference is not statistically significant ($p = 0.2175$).
\end{rqanswer}

\section{Discussion}
\label{sec:discussion}

\subsection{Comparison with the original study}

Table~\ref{tab:comparison} places the three languages side by side.
Two of them are replications performed for this report, and the Java column is taken from the original study.

\begin{table}[!t]
\caption{RQ1 results for the three languages. Java figures are from the original
study~\cite{wiratsin2025java}.}
\label{tab:comparison}
\centering
\begin{tabular}{lrrr}
\toprule
 & Java & Python & JavaScript \\
\midrule
Accepted answers                          & 874{,}438 & 840{,}132 & 1{,}144{,}185 \\
Code snippets in accepted answers         & 917{,}389 & 1{,}190{,}274 & 1{,}591{,}776 \\
Avg.\ snippets per accepted answer        & 1.05 & 1.42 & 1.39 \\
Answers edited at least once              & 140{,}840 & 346{,}535 & 447{,}379 \\
\quad as a percentage of accepted answers & 16.11\% & \textbf{41.25\%} & \textbf{39.10\%} \\
Avg.\ revisions per edited answer         & 2.82 & 2.78 & 2.68 \\
Median revisions per edited answer        & 2.0 & 2.0 & 2.0 \\
Maximum revisions                         & 53 & 111 & 68 \\
Code snippets across all revisions        & 283{,}838 & 1{,}640{,}080 & 1{,}998{,}016 \\
\midrule
Levenshtein distance, mean                & 88.87 & 49.66 & 48.66 \\
Levenshtein distance, maximum             & 21{,}158 & 28{,}500 & 20{,}690 \\
Levenshtein distance, std.\ dev.          & 359.44 & 268.53 & 228.44 \\
\bottomrule
\end{tabular}
\end{table}

\textbf{The premise of the original study holds, and holds more strongly.}
The share of accepted answers carrying at least one edit is 41.25\% for Python and 39.10\% for JavaScript, against 16.11\% for Java.
The supply of potential code improvements is therefore roughly two and a half times larger, in proportional terms, in both replication languages than in the original.
Combined with the higher number of code snippets per answer, this yields indexed corpora of 1{,}640{,}080 and 1{,}998{,}016 snippet revisions against Java's 283{,}838, six to seven times the size.
Whatever else differs, the motivation for the original study is not weaker outside Java; it is considerably stronger.

\textbf{Per-answer editing behaviour is almost invariant.}
The mean number of revisions per edited answer is 2.78 for Python, 2.68 for JavaScript and 2.82 for Java, and the median is exactly 2.0 in all three.
The three language communities differ substantially in \emph{how many} answers get edited, but barely at all in \emph{how much} an answer gets edited once someone starts.
That pattern is consistent with editing being driven by a common platform mechanism, the suggested-edit and review workflow, whose typical outcome is one or two accepted revisions, rather than by anything language-specific.

\textbf{Individual edits are smaller outside Java.}
The mean Levenshtein distance is 49.66 for Python and 48.66 for JavaScript, roughly half Java's 88.87.
Restricted to code snippets that actually changed, the means rise to 158.57 and 150.68, above the Java figure.
The safe statement is that Python and JavaScript edits are, if anything, more concentrated in a minority of code snippets, and that the two replication languages agree with each other far more closely than either agrees with Java.

\textbf{The popularity effect replicates in direction, and partially in significance.}
In both languages the number of matched answer edits rises monotonically from low- to medium- to high-popularity projects, on every one of the four measures we recorded.
This is the same direction the original study reported for both its Fixing Bug and Improving Code categories.
The Kruskal-Wallis test confirms the difference for Python ($p = 0.0018$), as it did for Java in both categories ($p = 0.0387$ and $p = 0.000016$), but not for JavaScript ($p = 0.2175$).

We do not think the JavaScript result should be read as a failure to replicate.
The original study tested across 10{,}673 projects; we tested across 100.
With 33 or 34 projects per group and a median of zero matched edits in most groups, the test has very little power to detect anything short of an overwhelming effect.
The trend is present and of the same shape as in Python, where the identical design does reach significance with roughly three times the counts.
The popularity effect is directionally confirmed in both languages and statistically confirmed in one.

\textbf{Why might popular projects match more edited answers?}
The original study framed its interpretation around code quality, expecting higher-popularity codebases to contain fewer suboptimal snippets and therefore to need fewer recommendations, and then found the opposite.
Our results reproduce that opposite.
The most straightforward explanation is size rather than quality: the original study reported that high-popularity Java projects average 112{,}896 lines against 18{,}453 for low-popularity ones, a sixfold difference, and a larger codebase simply offers more methods to match.
A second contributor is exposure, since widely used projects attract more contributors, more of whom may reach for a Stack Overflow snippet.
Our design cannot separate these, and we note it as a question the original study leaves open as well.

\subsection{Implications}

For researchers, the result strengthens the case for treating Stack Overflow answer revisions as a maintenance data source rather than as noise, and it does so in the two languages where the supply is largest.
The observation that roughly two thirds of code snippets are untouched by the edits to their answers is, to our knowledge, not reported elsewhere and has a practical consequence: a recommender built on this pipeline should filter on actual code change before ranking, or it will spend most of its candidate budget on code snippets with nothing to recommend.

For maintainers, the practical advice of the original study carries over unchanged, and applies to a larger body of code: when reusing a Stack Overflow snippet, record its provenance, because roughly two in five accepted Python and JavaScript answers are subsequently edited and there is currently no mechanism that will tell you when the answer you copied from has moved on.

\section{Threats to Validity}
\label{sec:threats}

\textbf{Construct validity.}
Our central assumption, inherited from the original study, is that a matched GitHub snippet represents an adoption of an earlier Stack Overflow answer that has since been improved.
Clone search establishes textual similarity between two snippets, not the direction of reuse or its origin.
A matched pair could equally reflect two independent implementations of a common task, code copied from GitHub to Stack Overflow, or reuse of a shared third-party source.
The original study mitigated this by manually inspecting every candidate pair; we do not.
Our counts should therefore be read as candidate matches, not as confirmed adoptions, and certainly not as confirmed improvements.
This is the most important limitation of this replication, and it is why our figures are comparable to the original study's 793 candidate pairs rather than to its 391 manually validated applicable recommendations.
It also means we can say nothing about \emph{what kind} of improvement the matched edits represent, since the categorisation into bug fixes and code improvements was a product of that manual step.

A second construct threat concerns the Levenshtein distance, which we use as a proxy for the magnitude of a code change.
Being a character-level measure, it is sensitive to superficial edits such as reformatting, whitespace and identifier renaming, and does not capture the semantic significance of a change.
We use it because the original study does, which keeps the comparison sound, but a small distance is not evidence of an unimportant change, nor a large one of an important change.

\textbf{Internal validity.}
Siamese+ was run on Python and JavaScript with the configuration tuned in the preparatory phase of the original study, which was optimised by Grid Search against a Java ground truth built from the Qualitas corpus, a collection of Java projects dating from 2013.
The parameters were therefore never tuned for, or validated against, Python or JavaScript code.
Values such as the minimum clone size of 6 lines and the n-gram sizes interact directly with a language's syntax and typical method length: Python is markedly more concise than Java and delimits blocks by indentation, so a 6-line threshold spans a larger unit of behaviour and may exclude valid matches, whereas JavaScript's callback and closure idioms produce shapes unlike those of Java methods.
Reusing the configuration keeps the replication faithful to the original methodology and makes the three languages directly comparable, but it may under- or over-report matches in either language.
The counts should be read as conservative and mutually comparable rather than as optimal for each language.
Retuning Siamese+ per language, following the same Grid Search procedure over a language-specific ground truth, is the obvious next step and is left as future work.

The clone search tool may also produce false positives and false negatives independently of its configuration, and the Stack Overflow data are drawn from the SOTorrent 2020-12-31 release, so they do not reflect the current state of the platform.

\textbf{Conclusion validity.}
The Kruskal-Wallis tests were applied to per-project counts that are extremely skewed, with a median of zero in five of the six groups.
The test is non-parametric and appropriate for such data, but with 33 or 34 projects per group its power is low, and the JavaScript result in particular should be read as inconclusive rather than as evidence of no effect.
The pairwise Mann-Whitney comparisons reported for context are uncorrected for multiple testing and are exploratory only.

\textbf{External validity.}
The most significant limitation is scale.
We searched 100 projects per language against the original study's 10{,}673, so our absolute counts are not comparable with the published Java figures and only the within-study distribution across groups should be compared.
Our groups are balanced by design, at 33, 33 and 34 projects, whereas the original study's were not, at 2{,}103, 5{,}073 and 3{,}497; the balanced design has the advantage that group size cannot itself drive the totals, but it also means our sample is not representative of the population of Python or JavaScript projects.

The popularity thresholds were inherited from the Java project population and applied unchanged, even though star and fork distributions differ across language ecosystems.
This keeps the three-way comparison meaningful but means the groups are not quartiles of the Python or JavaScript populations themselves.
Finally, our study, like the original, considers only accepted answers, so the findings may not extend to other kinds of answer, and it inherits the original's temporal limitation: the SOTorrent release predates the widespread adoption of generative AI coding assistants, so both editing behaviour and reuse dynamics on Stack Overflow may since have shifted.

\section{Conclusion}

We have replicated, on Python and JavaScript, an empirical study of Stack Overflow answer edits and their application to open-source code that was originally conducted on Java.
We analysed 840{,}132 accepted Python answers and 1{,}144{,}185 accepted JavaScript answers from SOTorrent, using the same extraction pipeline, the same clone search tool and the same project popularity criteria as the original study.

The central findings generalise.
In both languages the proportion of accepted answers that have been edited is far higher than in Java, 41.25\% and 39.10\% against 16.11\%, so the supply of potential code improvements is substantially larger outside Java than within it.
The number of revisions per edited answer, by contrast, is almost invariant across the three languages at 2.78, 2.68 and 2.82, with a median of exactly 2.0 in every case, suggesting that the editing mechanism itself is a property of the platform rather than of any one community.
Searching 100 GitHub projects per language, Siamese+ matched 1{,}080 edited Python code snippets and 449 edited JavaScript ones, and in both languages the counts increase monotonically from low- to medium- to high-popularity projects, reproducing the distribution reported for Java.
That difference is statistically significant for Python but not for JavaScript, where the small number of projects per group leaves the test with little power.

Two observations fall outside the original study's scope and are worth carrying forward.
First, roughly two thirds of the code snippets in edited answers are byte-identical between their first and latest revisions, in both languages, so an answer being edited is a weak predictor of its code having changed; a recommender built on this pipeline should filter on actual code change early.
Second, the two replication languages agree with each other far more closely than either agrees with Java on every distributional measure we collected, which is what one would hope to see if the underlying phenomenon is real.

The clearest directions for future work follow from this report's limitations.
Retuning Siamese+ against language-specific ground truths would establish how much of the difference in match counts is attributable to a Java-optimised configuration.
Extending the search to a project sample comparable in size to the original study's would give the popularity analysis the statistical power that the JavaScript result currently lacks.
Manual validation, or an automated classifier trained on the original study's labelled Java data, would turn our candidate matches into applicability rates directly comparable with the 49.30\% reported for Java, and would open the way to the pull-request evaluation that closed the original study.

\section*{Data Availability}

The replication package for this report is publicly available on
Zenodo~\cite{replication2026package} at \url{https://doi.org/10.5281/zenodo.21900949}.
It contains the analysis scripts, the extracted Stack Overflow code snippet revisions, and the derived datasets behind the results reported in Sections~\ref{sec:results-python} and~\ref{sec:results-javascript}.
The SOTorrent 2020-12-31 release~\cite{baltes2018sotorrent} from which the snippets were extracted, and Siamese+~\cite{wiratsin2025java} with the Python and JavaScript support used here, are available separately from their own sources.

\bibliographystyle{IEEEtran}
\bibliography{references}

\end{document}